\documentclass[11pt]{article}
\usepackage{latexsym}
 \usepackage[dvips]{graphics}
\usepackage{amsmath}  
\usepackage[dvips]{graphics}
\usepackage{latexsym}
\usepackage{amscd}     \usepackage{amsxtra}
\usepackage{upref}     \usepackage{amsthm}
\usepackage{amssymb}   
\usepackage{amsfonts}
\begin{document}

\title{{\bf  On the Pre-metric Formulation and Nonlinearization of
Charge-free Electrodynamics}}

\author{
{\bf Stoil Donev}\footnote{e-mail:sdonev@inrne.bas.bg},
{\bf Maria Tashkova}, \\
Institute for Nuclear Research and Nuclear Energy,\\ Bulg.Acad.Sci., 1784 Sofia,
blvd.Tzarigradsko chaussee 72, Bulgaria\\}
\date{}
\maketitle

\begin{abstract}
This paper presents a coordinate free pre-metric formulation
of charge free Maxwell-Minkowski  electrodynamics, and of the developed by the
authors non-linear Extended Electrodynamics. First we introduce some formal
relations from multilinear algebra and differential geometry to be used
further. Then we recall and appropriately modify the existing pre-metric
formulation of linear charge free electrodynamics in pre-relativistic and
relativistic forms as preparation to turn to corresponding pre-metric
nonlinearization. After some preliminary examples and notes on
nonlinearization, we motivate our view for existence and explicit formulation
of time stable subsystems of the physical field objects considered. Section 5
presents the formal results of our approach on the pre-metric nonlinear
formulations in static case, in time-dependent case, and in space-time
formulation. In the Conclusion we give our general view on "why and how to
nonlinearize". The Appendix gives a possible formal extension of our aproach to
many subsystem field objects.

\end{abstract}

{\it keywords}:
extended electrodynamics, electromagnetic field objects,
pre-metric, nonlinearization.

\section{Introduction}
We start with some general remarks concerning our view on physical systems and
the necessary physical motivations when a theoretical model is meant to be
appropriately constructed. Appropriate here means that, the description of
structure and behavior must be based on the natural assumption that physical
world demonstrates itself through relatively stable objects, being able to
survive, or to be destroyed, during mutual interactions.  Our knowledge about
physical world comes namely from this mutual interaction, no matter weather
this interaction is direct, or indirect. We should not equalize the concepts of
{\it explanation} and {\it description}, our view is that theoretical physics
must {\it explain} any description of natural objects and processes. So,
physical motivations should lead to explanation of the description model built.
Every built model must give: first, corresponding mathematical structure of the
physical system studied, which includes {\it constituents $+$ subsystems},
second, mathematical images of the physical characteristics by means of which
the system considered demonstrates {\it existence} and {\it participation}
among the other systems. For example, it is one thing to model natural flows by
vector fields, and another thing to calculate the exchanged energy-momentum
between these flows. We may conclude that mathematical models should
rather give equations of balance of corresponding quantities, than equations
that equalize the mathematical images of the interacting systems. For example,
Poynting's equation
$div(\mathbf{E}\times\mathbf{B})=-
\frac{1}{2c}\frac{\partial }{\partial t}(\mathbf{E}^2+\mathbf{B}^2)$
is a balance relation, while Maxwell equation
$\frac1c\frac{\partial \mathbf{E}}{\partial t}=
curl\,\mathbf{B}$ equalizes differentiated mathematical images.

 Any physical system has spatial structure and shows definite stability
properties, so, it can support its existence and compensate in definite degree
the external disturbances through appropriate shape changes and kinematical
behavior without losing identity. Shortly speaking, its time existence is a
dynamical process being strongly connected with various and continuous internal
and external stress-energy-momentum exchange processes. All these processes are
real phenomena and any attempt for their description should be based on
appropriate mathematical structures. Of course, all real changes during
existence of a physical system also must be formally identified with
appropriate mathematical objects. Clearly, if we are going to describe formally
a real physical system together with all admissible changes, the corresponding
mathematical images of the system, including its recognizable subsystems and
all real changes, must have {\it tensor} nature in order not be possible to be
removed by coordinate transformations.

All these moments suggest to pay due respect to modern differential geometry as
appropriate mathematical language. In fact, the more than a century intensive
interaction between differential geometry and theoretical and mathematical
physics proved to be exclusively useful, suggestive and creative process.

The plan of the paper is the following.

First we introduce some formal relations from multilinear algebra and
differential geometry to be used further. Then in sections 2-3 we start the
pre-metric approach in the linear charge free electrodynamics in
pre-relativistic and relativistic forms as preparation to turn to
corresponding pre-metric nonlinearization. Section 4 gives some preliminary
examples and notes and on nonlinearization, and motivates our view for
existence and explicitly formulation of time stable subsystems of the physical
objects considered.  Section 5 presents the pre-metric formulation in the
static case, in time-dependent case, and in space-time formulation. In the
Conclusion we give our general view on "why and how to nonlinearize".

In the next section we are going to introduce the objects and relations to be
used further in the paper. Our basic mathematical object will be an oriented
manifold $(M,\omega)$ with appropriate dimension. The orientation defined by
$\omega$ will allow to compute integral characteristics of the objects
considered on one hand, and to make use of the Poincare isomorphisms between
$p$-vector fields and $(dim\,M-p)$-exterior forms. The two basic
operators to use will be, of course, the interior product $i_{T}\alpha$, where
$T$ is a $p$-vector, $\alpha$ is a $q$-form, $p\leqq q$,
and the exterior derivative
$\mathbf{d}: \Lambda^p(M)\rightarrow\Lambda^{p+1}(M)$.

\section{Some formal relations}

Let  $E$ and $E^*$ be two dual real finite dimensional vector spaces.
The duality between $E$ and $E^*$ allows to distinguish the following  well
known (anti)derivation. Let $h\in E$, then we obtain the derivation $i(h)$, or
$i_h$, in $\Lambda(E^*)$ of degree $(-1)$ (sometimes called
substitution/contraction/insertion operator, interior product, algebraic flow)
 according to [1]:
$$
i(h)(x^{*1}\wedge\dots\wedge
x^{*p})=\sum_{i=1}^{p}(-1)^{(i-1)}\langle x^{*i},h\rangle
x^{*1}\wedge\dots\wedge\hat{x^{*i}}\wedge
\dots\wedge x^{*p}.
$$
Clearly, if $u^*\in \Lambda^p(E^*)$ and $v^*\in\Lambda(E^*)$ then
$$
i(h)(u^*\wedge v^*)=(i(h)u^*)\wedge v^*+(-1)^pu^*\wedge i(h)v^*.
$$
Also, we get
$$
i(h)u^*(x_1,\dots,x_{p-1})=u^*(h,x_1,\dots,x_{p-1}), \
$$
$$
i(x)\circ i(y)=-i(y)\circ i(x).
$$

This antiderivation is extended to a mapping $i(h_1\wedge\dots\wedge h_p):
\Lambda^m(E^*)\rightarrow\Lambda^{(m-p)}(E^*)$, $m\geqq p$,
according to
$$
i(h_1\wedge h_2\wedge\dots\wedge h_p)u^*=i(h_p)\circ\dots\circ i(h_1)\,u^*.
$$
Note that this extended mapping is not an antiderivation, except for $p=1$.

Clearly, this interior product may be used in "opposite direction", i.e.,
if $u^*$ is 1-form, then
$$
i(u^*)(h_1\wedge h_2\wedge ... \wedge h_p)=<u^*,h_1>(h_2\wedge ... \wedge
h_n)- <u^*,h_2>(h_1\wedge h_3\wedge ... \wedge h_p)+ ...
$$
$$
+(-1)^{(p-1)}
<u^*,h_p>(h_1\wedge h_2\wedge ... \wedge h_{p-1}).
$$


This mapping is extended to multivectors and exterior forms which are linear
combinations: if $\Psi=\Psi_1+\Psi_2+...$ is an arbitrary multivector on $E$ and
$\Phi=\Phi^1+\Phi^2+...$ is an arbitrary exterior form on $E^*$ then
$i_{\Psi}\Phi$ is defined as extension by linearity, e.g.,
$$
i(\Psi_1+\Psi_2)(\Phi^1+\Phi^2)
=i(\Psi_1)\Phi^1+i(\Psi_1)\Phi^2+i(\Psi_2)\Phi^1+i(\Psi_2)\Phi^2.
$$

The above extension of the interior product allows to extend the Lie derivative
of a differential form $\Phi$ along a vector field $X$ to a derivative of
$\Phi$ along a multivector field $T$ [2], according to
\begin{equation}
\mathcal{L}_T(\Phi)=
d\circ i_T\Phi- (-1)^{deg(T)}i_T\circ d\Phi.
\end{equation}
If $\mathcal{L}_T(\Phi)=0$ this extension allows to consider $T$ as a local
symmetry of $\Phi$.

The specialization of these interior products for the cases $i(\Phi)\omega^*$ and
$i(\beta)\omega$, where $\Phi$ is a multivector,
$\omega^*=\varepsilon^1\wedge\varepsilon^2\wedge ... \wedge\varepsilon^n$ is a
volume form in $E^*$, $\beta$ is a $p$-form, and
$\omega=e_1\wedge e_2\wedge ... \wedge e_n$ is the volume form in $E$, {\it
dual} to $\omega^*$ :$<\omega^*,\omega>=1$, is well known [1] and used in
multilinear algebra and differential topology under the name of Poincare
isomorphisms, or Poincare dualities. In fact,
we note that the two spaces $\Lambda^p(E)\otimes\Lambda^n(E^*)$
and $\Lambda^{n-p}(E^*)$
have the same dimension, so, every nonzero $\omega\in\Lambda^n(E^*)$
generates isomorphism $D^{p}$, between these two spaces according to
$(u,\omega^*)\rightarrow i(u)\omega^*$, where $u\in\Lambda^p(E)$ is a $p$-vector
over $E$. In particular, if $\{e_i\}$ and $\{\varepsilon^j\}$ are dual bases,
the corresponding basis elements
$$
e_{\nu_1}\wedge\dots\wedge e_{\nu_p},\ \ \ \nu_1<\nu_2<...<\nu_p ,
$$
and
$$\varepsilon^{\nu_{p+1}}\wedge\dots\wedge\varepsilon^{\nu_n} , \ \ \
\nu_{p+1}<\nu_{p+2}<...<\nu_n,
$$ are connected according to
$$
D^{p}(e_{\nu_1}\wedge\dots\wedge e_{\nu_p})=
(-1)^{\sigma}\varepsilon^{\nu_{p+1}}\wedge\dots\wedge\varepsilon^{\nu_n},
$$
where $\sigma=\sum_{i=1}^p(\nu_i-i)$. Also,
$$
D_{p}(\varepsilon^{\nu_1}\wedge\dots\wedge \varepsilon^{\nu_p})=
(-1)^{\sigma}e_{\nu_{p+1}}\wedge\dots\wedge e_{\nu_n},
$$
$$
D_{p}(\varepsilon^{\nu_{p+1}}\wedge\dots\wedge \varepsilon^{\nu_n})=
(-1)^{p(n-p)+\sigma}e_{\nu_1}\wedge\dots\wedge e_{\nu_p}.
$$
Clearly, we have
$$
i(e_{\nu_1}\wedge\dots\wedge e_{\nu_p})
D^{p}(e_{\nu_1}\wedge\dots\wedge e_{\nu_p})=0.
$$
Also, we note that in this way every subspace $V^p\subset E$ leads to
defining three other spaces:
$$
(V^p)^*\subset E^*; \ \ D^p(V^p)\subset E^*; \ \
 D_p((V^p)^*)\subset E \ ,
$$
where
$$
E=V^p\oplus(D^p(V^p))^* \ ; \ E^*=(V^p)^*\oplus D^p(V^p) .
$$
We remind that these isomorphisms depend on the chosen
element $\omega\in\Lambda^n(E^*)$, but further in this section we shall omit
writing $\omega$ for clarity.

 These last two formulas allow to make use of any
isomorphism between $E$ and $E^*$ for defining isomorphisms
$\Lambda^p(E)\cong\Lambda^{n-p}(E)$, and
$\Lambda^p(E^*)\cong\Lambda^{n-p}(E^*)$.

For these isomorphisms and their duals
$$
D^p:\Lambda^p(E)\rightarrow\Lambda^{n-p}(E^*), \ \
(D^p)^*: \Lambda^{n-p}(E)\rightarrow\Lambda^p(E^*)
$$
$$
D_p:\Lambda^p(E^*)\rightarrow\Lambda^{n-p}(E), \ \
(D_p)^*: \Lambda^{n-p}(E^*)\rightarrow\Lambda^p(E) ,
$$
the following relations also hold:
$$ (D_p)^*=(D^p)^{-1}=(-1)^{p(n-p)}D_{n-p} \ ; \ \
(D^p)^*=(D_p)^{-1}=(-1)^{p(n-p)}D^{n-p} \ ;
$$
$$ D_{n-p}\circ D^p=(-1)^{p(n-p)}id , \ \
D^{n-p}\circ D_p=(-1)^{p(n-p)}id ,
$$
where $id$ denotes the corresponding identity map. So,
up to a sign factor, $D^p$ and $D_{n-p}$ are inverse linear isomorphisms.
When $T\in\Lambda(E)$ and $\Psi\in\Lambda(E^*)$
are represented by exterior products, for example:
$T=u\wedge v$, $\Psi=\alpha\wedge\beta$, and their degrees are not given,
then we replace $D^{p}$ by $\mathfrak{D}$ and
$D_{p}$ by $\mathbb{D}$ , and the following relations hold
$$
\mathfrak{D}(T)=i_{v}\circ i_{u}\omega,
\ \ \ \mathbb{D}(\Psi)=i_{\beta}\circ i_{\alpha}\omega^{*}.
$$
These relations will be used further in Sec.2.4.

 We extend now these insertion operators in different direction. Let
$E_1$ and $E_2$ be two real vector spaces with corresponding bases
$\{e_i\}$,$i=1,2,...,dimE_1$ and $\{k_j\}$,$j=1,2,...,dimE_2$,
$T=\mathfrak{t}^i\otimes e_i$
be a $E_1$-valued $q$-vector, $\Phi=\alpha^j\otimes k_j$ be a $E_2$-valued
$p$-form with $q\leq p$, and $\varphi:E_1\times E_2\rightarrow F$ be a bilinear
map into the vector space $F$. Now we define
$i^{\varphi}_T\Phi\in\Lambda^{p-q}(M,F)$:

\begin{equation}
 i^{\varphi}_T\Phi= i_{\mathfrak{t}^i}\alpha^j\otimes\varphi(e_i,k_j), \ \ \
i=1,2,...,dim(E_1), \ j=1,2,...,dim(E_2).
\end{equation}
Also, if $T_1, T_2$ are two multivectors and $\alpha,\beta$ are two
forms then $(i\otimes i)_{T_1\otimes T_2}(\alpha\otimes \beta)$ is defined by
$$
(i\otimes i)_{T_1\otimes T_2}(\alpha\otimes \beta)=
i_{T_1}\alpha\otimes i_{T_2}\beta.
$$

We can define now the $\varphi$-extended Lie derivative. Let $M$ be a
$n$-dimensional manifold, $\Phi$ be a $E_1$-valued differential $p-$form on
$M$, $T\in\mathfrak{X}^q(M,E_2)$ be a $E_2$-valued $q$-multivector field on
$M$, with $q\leq p$ and $\varphi:E_1\times E_2\rightarrow F$ be a bilinear map.
The $\varphi$-extended Lie derivative

$$
\mathcal{L}^{\varphi}_T:
\Lambda^p(M,E_1)\times\mathfrak{X}^q(M,E_2)\rightarrow\Lambda^{p-q+1}(M,F)
$$
is defined as follows [21, p.53]:
\begin{equation}
\mathcal{L}^{\varphi}_T(\Phi)=
\mathbf{d}\circ i^{\varphi}_T\Phi-
(-1)^{deg(T).deg(\mathbf{d})}i^{\varphi}_T\circ\mathbf{d}\Phi,
\end{equation}
where $\mathbf{d}$ is the exterior derivative on $M$, so, $deg(\mathbf{d})=1$.
This definition suggests to consider the tensor field $T$ as a {\it local}
$\varphi$-symmetry of the differential form $\Phi$ when
$\mathcal{L}^{\varphi}_T(\Phi)=0$.

\section{Pre-metric formulation in the linear case}

Since we are not going to use metric tensors in the cases considered, for more
information, including history, analysis and citations, on the so-called {\it
pre-metric} formulation of electrodynamics, not only in the charge free case,
the reader may find in the 2-volume e-book by D.A.Delphenich [3].

\subsection{Pre-relativistic approach}
Let $X$ be a vector field on $(\mathbb{R}^3,\omega)$, where
$\omega =dx\wedge dy\wedge dz$ is the usual volume 3-form on $\mathbb{R}^3$.
The usual interaction partners of any vector field  X are the differential forms
$\alpha\in \Lambda(\mathbb{R}^3)$ on the same manifold since the generated by
$X$ flow $i_{X}\alpha$ across $\alpha$ may change $\alpha$, i.e.,
$i_{X}\alpha$ may be not zero: $i_{X}\alpha\neq 0$. In searching for such a
differential form partner, our vector field $X$ finds only one that is
specially indicated, the volume 3-form $\omega$. Now the following question
arizes: does this flow $i_{X}\omega\neq 0$ change from point to point? Formally
this means whether the Lie derivative $L_{X}\omega$ change from point to point
along $X$, or not. For this Lie derivative we obtain

$$
L_{X}\omega=\mathbf{d}\,i_{X}\omega+i_{X}\mathbf{d}\omega=\mathbf{d}\,i_{X}\omega =
\mathbf{d}\left[X^1dy\wedge dz-X^2dx\wedge dz+X^3dx\wedge dy\right]
$$
since $\mathbf{d}\omega=0$ by dimensional reasons. So, if this flow does not
change from point to point, the 2-form $i_{X}\omega$ must be closed:
$\mathbf{d}\,i_{X}\omega=0$. As is easily verified, this means that the vector
field $X$ has zero divergence:

$$
\mathbf{d}\,i_{X}\omega= \left[\frac{\partial X^1}{\partial x}+
\frac{\partial X^2}{\partial y}+ \frac{\partial X^3}{\partial z}\right]dx\wedge
dy\wedge dz= (div\,X).\omega=0\ \ \ \rightarrow \ \ \ div\,X=0.
$$

We specially note that the 2-form $i_{X}\omega$ is {\it closed} :
$\mathbf{d}\,i_{X}\omega=0$. Therefore, there exist a class of 1-forms
$\alpha+df$ such that, locally, we may write
$i_{X}\omega=\mathbf{d}(\alpha+df)$. We conclude:

\vskip 0.3cm
{\it Every
divergence-free vector field on $(\mathbb{R}^3,\omega)$ is naturally connected
with some differential 1-forms}. \vskip 0.3cm Recalling now the two Maxwell
equations $div\,\mathbf{E}=0, div\,\mathbf{B}=0$ we conclude also, that these
two vector fields $(\mathbf{E},\mathbf{B})$ are {\it intrinsically} connected
with corresponding differential 1-forms, and this connection does NOT require
any metric tensor.

\vskip 0.3cm
However, if the flows generated by $\mathbf{E}$ and $\mathbf{B}$ admit
{\it time-change} although each of them does not change the volume form
$\omega$? This question suggests to look more carefully for other possible ways
for mutual influence of the two flows generated by $\mathbf{E}$ and
$\mathbf{B}$.
  \vskip 0.3cm
According to the above notations the two  {\it lineary independent}
vector fields $(\mathbf{E},\mathbf{B})$ appear together with their
corresponding 2-forms $i_{\mathbf{E}}\omega$ and $i_{\mathbf{B}}\omega$.
Moreover, the equations $div\,\mathbf{E}=0, div\,\mathbf{B}=0$ require  these
two 2-forms $(i_{\mathbf{E}}\omega, \, i_{\mathbf{B}}\omega)$ to be closed, and
so - locally exact. We conclude that there {\it must exist} two 1-forms
$\alpha$ and $\beta$ satisfying the equations
$$
i_{\mathbf{E}}\omega=\mathbf{d}\alpha , \ \ \
i_{\mathbf{B}}\omega=\mathbf{d}\beta, \ \ \  i_{\mathbf{E}}\mathbf{d}\alpha=0,
\ \ \ i_{\mathbf{B}}\mathbf{d}\beta=0.
$$

Another suggesting moment in this direction is the seriously ised {\it "curl"}
operator in Maxwell equations. Making use of the mentioned Poincare
isomorphisms between $p$-forms and $(n-p)$-vectors on oriented manifolds,
it is easily found that this {\it "curl"} operator is
strongly connected to the exterior derivative operator $\mathbf{d}$, acting in
the graded algebra of differential forms. Explicitly, in the case of our
manifold $(\mathbb{R}^3,\omega)$, the restriction $\mathbb{D}$ of the Poincare
isomorphism to differential 2-forms of the kind $\mathbf{d}\alpha$, where
$\alpha$ is arbitrary 1-form, maps $\mathbf{d}\alpha$ to a unique vector field
of the kind $curl(\mathbf{V})$:
$$
\mathbb{D}(\mathbf{d}\alpha)=curl(\mathbf{V}), \ \
\text{or}, \ \ \mathfrak{D}(curl(\mathbf{V}))=\mathbf{d}\alpha ,
$$
and the components
of $\mathbf{d}\alpha$ in the basis $(dx\wedge dy,dx\wedge dz,dy\wedge dz)$
coincide with the components of $curl(\mathbf{V})$ in the basis
$\left(\frac{\partial}{\partial x},\frac{\partial}{\partial y},
\frac{\partial}{\partial z}\right)$.
Of course, most easily this can be achieved if the
components of $\alpha$ and the components of $\mathbf{V}$ are the same in the
corresponding dual bases $(dx,dy,dz)\leftrightarrow
\left(\frac{\partial}{\partial x},\frac{\partial}{\partial y},
\frac{\partial}{\partial z}\right)$.

Also, Poincare isomorphism generates two bivectors $\bar{H}$ and $\bar{K}$
according to
$$
\alpha=i_{\bar{H}}\omega , \ \ \ \beta=i_{\bar{K}}\omega.
$$

 All this  suggests to look for realizable
$\mathbf{E}\leftrightarrow\mathbf{B}$ {\it time}-interaction making use of
$\alpha$ and $\beta$.
\vskip 0.3cm
In Maxwell theory time is introduced as {\it
external} to the coordinates $(x,y,z)$ parameter, denoted by $t$,
so, the time
differentiation of tensor fields on $\mathbb{R}^3$ does NOT change the tensor
nature of the object. It seems that  $i_{\mathbf{E}}\omega$ and
$i_{\mathbf{B}}\omega$ may have a chance to find their corresponding {\it
"curl"} partners paying appropriate respect to their own time derivatives. This
suggests to answer positively to the question "do available non-zero time
derivatives of $(\mathbf{E},\mathbf{B})$ find way for realization through the
1-form partners". As is well known, classical charge free electrodynamics, as
represented by Maxwell equations, gives a positive answer to this question. In
our approach it looks as follows.

The closed nature of $i_{\mathbf{E}}\omega$ and $i_{\mathbf{B}}\omega$ and
the above consideration and notations allow in the corresponding bases
$\left(\frac{\partial}{\partial x},\frac{\partial}{\partial y},
\frac{\partial}{\partial z}\right)$ and
$(dx\wedge dy,dx\wedge dz,dy\wedge dz)$ to write down
$$
\mathfrak{D}(curl(\mathbf{E}))=\mathbf{d}\alpha,  \ \ \text{or}, \ \
\mathbb{D}(\mathbf{d}\alpha)=curl(\mathbf{E}),
$$ and
$$
\mathfrak{D}(curl(\mathbf{B}))=\mathbf{d}\beta, \ \ \text{or}, \ \
\mathbb{D}(\mathbf{d}\beta)=curl(\mathbf{B}).
$$

 Now, in order to model mutual
and {\it time-realizable} interaction between $\mathbf{E}$ and $\mathbf{B}$
making help of $\alpha$ and $\beta$, {\it Maxwell consistency condition}
requires
\begin{equation} \frac1c\frac{\partial }{\partial
t}i_{\mathbf{E}}\omega=\mathbf{d}\beta , \ \ \ \frac1c\frac{\partial }{\partial
t}i_{\mathbf{B}}\omega=-\mathbf{d}\alpha ,
\end{equation}
where $c$ denotes an
invariant constant velocity, characterizing the translational propagation of
the electromagnetic field object considered. Hence, equations (4) formally
introduce a kind of {\it cross-connection} between the {\it divergence-free
nature of} $(\mathbf{E},\mathbf{B})$ and the {\it external} nature of their
possible time dependence.

So, the content of this positive answer physically means that time derivatives
of $(\mathbf{E},\mathbf{B})$ {\it interchange} their 1-form partners: a time
change of $\mathbf{E}$ requires appropriate local spatial change of the 1-form
partner $\beta$ of $\mathbf{B}$, and a time change of $\mathbf{B}$ requires
appropriate local spatial change of the 1-form partner $\alpha$ of
$\mathbf{E}$. We note that, from formal viewpoint, this partnership makes use
of the {\it invariant} operator {\bf exterior derivative}, so the partnership
is {\bf real}, and the introduced cross-connection of this reality with the
corresponding time derivatives allows to consider the time dependent
electromagnetic field objects as {\bf real} ones.

Recall now the concept of local {\it helicity} generated by any 1-form $\eta$
on $\mathbb{R}^3$, it is defined as the 3-form $\eta\wedge\mathbf{d}\eta$ and
carries information about the rotational properties of the flow generated by
a divergence-free vector field $X$ that is euclidean metric image of $\eta:
\eta_i=g_{ij}X^j$, or $X^i=g^{ij}\eta_j$, where $g$ denotes the euclidean
metric tensor on $\mathbb{R}^3$. In fact,
$\eta\wedge\mathbf{d}\eta=g(X,curl\,X).\omega$.

The above equations (4) determine time-partnership respectively between
$(\mathbf{E},\beta)$ and between $(\mathbf{B},\alpha)$. This suggests to show
appropriate respect to the corresponding {\it cross helicities}, given by
$\alpha\wedge\mathbf{d}\beta$ and $\beta\wedge\mathbf{d}\alpha$. From classical
vector analysis we have the relation

$$
L_{\mathbf{E}\times\mathbf{B}}\omega=
\left[\mathbf{B}.curl(\mathbf{E})-\mathbf{E}.curl(\mathbf{B})\right]\omega=
div(\mathbf{E}\times\mathbf{B}).\omega .
$$

The corresponding Poynting relation
in Maxwell theory suggests now the following. Since our field object will
{\it propagate} in the 3-space, then at the points where the field functions
are not zero at the moment $t$, generally speaking, there must be expected time
changes of the {\it important local characteristics} of the object at these
points. Trying to present these changes without making use of a metric, we can
make use of the above introduced cross helicities as follows:
 $$
\beta\wedge\mathbf{d}\alpha-\alpha\wedge\mathbf{d}\beta
=-\frac{\partial}{\partial \xi}
\left(\frac{<\alpha,\mathbf{E}>+<\beta,\mathbf{B}>}{2}\right).\,\omega, \
\xi=ct .
 $$

We form now the following $(1,1)$ tensor:
$$
\mathbb{T}=\alpha\otimes\mathbf{E}+\beta\otimes\mathbf{B}
-\frac12\left(<\alpha,\mathbf{E}>+
<\beta,\mathbf{B}>\right)id_{T\mathbb{R}^3}.
$$
Obviously, this tensor represents a {\it metric-free} form of Maxwell stress
tensor.

A more careful look at $\mathbb{T}$ naturally sets the question "weather the
two vector fields $\mathbf{E}$ and $\mathbf{B}$, which may be
considered as {\it constituents}
of $\mathbb{T}$, define eigen directions of $\mathbb{T}$ at every point". The
corresponding to this question equations read:
$$
i_{\mathbf{E}}\mathbb{T}=\lambda_1.\mathbf{E}, \ \ \
i_{\mathbf{B}}\mathbb{T}=\lambda_2.\mathbf{B}.
$$
Since $\mathbf{E}$ and $\mathbf{B}$ are lineary independent, it is
elementary to see that these two relations hold only if
$$
<\alpha,\mathbf{B}>=0, \ \ <\beta,\mathbf{E}>=0.
$$
So, we have two more algebraic equations connecting
$(\mathbf{E},\mathbf{B})$ with $(\alpha,\beta)$.

Finally, the mutually dependent vector fields
$(\mathbf{E},\mathbf{B})$ define the bi-vector
$\mathbf{E}\wedge\mathbf{B}$,
and a natural algebraic characteristic of the assumed dependence between
$(\mathbf{E},\mathbf{B})$ and $(\alpha,\beta)$ appears to be the flow of
of $\mathbf{E}\wedge\mathbf{B}$ across the 2-form $\alpha\wedge\beta$.
In view of the above last relations for this flow we obtain
$$
i_{\mathbf{E}\wedge\mathbf{B}}(\alpha\wedge\beta)=
i_{\mathbf{B}}\circ\,i_{\mathbf{E}}(\alpha\wedge\beta)=
<\alpha,\mathbf{E}><\beta,\mathbf{B}>-
<\alpha,\mathbf{B}><\beta,\mathbf{E}>=<\alpha,\mathbf{E}><\beta,\mathbf{B}>.
$$
We may assume now the following way to relate this nonlinear flow with the linear
flow of $\mathbf{E}$ across $\alpha$, given by $<\alpha,\mathbf{E}>$,
and of $\mathbf{B}$ across $\beta$, given by $<\beta,\mathbf{B}>$:

$$
i_{\mathbf{E}\wedge\mathbf{B}}(\alpha\wedge\beta)=
\Big[\frac{<\alpha,\mathbf{E}>+<\beta,\mathbf{B}>}{2}\Big]^2.
$$
From these last two algebraic relations it will follow
$$
<\alpha,\mathbf{E}>^2-2<\alpha,\mathbf{E}>.<\beta,\mathbf{B}>+
<\beta,\mathbf{B}>^2=(<\alpha,\mathbf{E}>-<\beta,\mathbf{B}>)^2=0,
$$
i.e., $<\alpha,\mathbf{E}>=<\beta,\mathbf{B}>$.

\vskip 0.4cm

{\bf Remark}.

{\it Considering the flow of  $\mathbf{E}\wedge\mathbf{B}$ across $\omega$ we get
the 1-form
$$
\theta=i_{\mathbf{E}\wedge\mathbf{B}}\omega.
$$
Since $i_{\mathbf{E}\wedge\mathbf{B}}=i_{\mathbf{B}}\circ\,i_{\mathbf{E}}$, we
get $i_{\mathbf{E}}\theta=i_{\mathbf{B}}\theta=0$, therefore, this 1-form
$\theta$ may carry information about the integrability of the 2-dimensional
distribution $\Gamma$ on $\mathbb{R}^3$ defined by $\mathbf{E}$ and
$\mathbf{B}$. It seems reasonable to admit that in some cases $\Gamma$ may be
integrable which formally means that $\theta$ must satisfy the equation $$
\mathbf{d}\theta\wedge \theta=0,
$$
for example, when time stable and spatially finite solutions are under
consideration. In the considered linear case it is hard to believe that such
solutions exist at all, but let it be in sight}.

\vskip 0.4cm In this way we
have $11$ equations for the $12$ components of
$\mathbf{E},\mathbf{B},\alpha,\beta$. They read:

\begin{equation}
L_{\mathbf{E}}\omega=0,
\ \ L_{\mathbf{B}}\omega=0, \ \
\frac{\partial }{\partial \xi}i_{\mathbf{E}}\omega=\mathbf{d}\beta , \ \
\frac{\partial }{\partial \xi}i_{\mathbf{B}}\omega=-\mathbf{d}\alpha, \ \
\end{equation}
\begin{equation}
i_{\mathbf{E}}\mathbb{T}=\lambda_1.\mathbf{E}, \ \ \
i_{\mathbf{B}}\mathbb{T}=\lambda_2.\mathbf{B}, \ \ \
<\alpha,\mathbf{E}>=<\beta,\mathbf{B}>, \ \ \xi=ct.
\end{equation}

These equations do {\it not} use any metric, so, this approach may represent the
{\it nonrelativistic} view on so called {\it pre-metric} formulation of charge
free Maxwell equations.

\vskip 0.4cm
\subsection{Relativistic approach}
The relativistic approach to electrodynamic phenomena seriously uses the
assumption that electromagnetic field objects propagate in free space with
constant speed {\it c}, where "constant" means {\it "the same with respect to
any reference frame"}, i.e., with respect to any free mass body. This brought the
necessity to introduce the notion that time-measuring depends on the reference
frame with respect to which the time-measuring system does not move.
In this way the fourth coordinate $\xi=ct$, $t$ is the measured time interval,
was introduced, the 3-space $\mathbb{R}^3$ was extended to $\mathbb{R}^4$, and
the 3-space volume form $\omega_{3}=dx\wedge dy\wedge dz$ was extended to
4-spacetime volume form $\omega$:
$$
\omega=\omega_{3}\wedge d\xi=dx\wedge dy\wedge dz\wedge d\xi,
$$
where $d\xi=cdt$.

The above mentioned {\it Maxwell consistency condition} suggests the following
{\it extension} of the two 2-forms $i_{\mathbf{E}}\omega_{3}$ and
$i_{\mathbf{B}}\omega_{3}$ to two spacetime 2-forms:
$$
F=i_{\mathbf{B}}\omega_{3}+\alpha\wedge d\xi= B^3\,dx\wedge dy-B^2\,dx\wedge
dz+B^1\,dy\wedge dz+\alpha_1\,dx\wedge d\xi+\alpha_2\,dy\wedge
d\xi+\alpha_3\,dz\wedge d\xi,
$$
$$ G=i_{\mathbf{E}}\omega_{3}-\beta\wedge
d\xi= E^3\,dx\wedge dy-E^2\,dx\wedge dz+E^1\,dy\wedge dz-\beta_1\,dx\wedge
d\xi-\beta_2\,dy\wedge d\xi-\beta_3\,dz\wedge d\xi,
$$
where $\mathbf{E}^i=\alpha_i$, and $\mathbf{B}^i=\beta_i, i=1,2,3$.
Making use now of the new volume 4-form $\omega$, the corresponding bi-vector
fields $\bar{F}$ and $\bar{G}$ are defined by
$$
F=i_{\bar{G}}\,\omega, \ \ \ G=-i_{\bar{F}}\,\omega.
$$
Explicitly,
$$
\bar{F}=\beta_3\frac{\partial }{\partial x}\wedge\frac{\partial }{\partial y}-
\beta_2\frac{\partial }{\partial x}\wedge\frac{\partial }{\partial z}+
\beta_1\frac{\partial }{\partial y}\wedge\frac{\partial }{\partial z}-
E^1\frac{\partial }{\partial x}\wedge\frac{\partial }{\partial \xi}-
E^2\frac{\partial }{\partial y}\wedge\frac{\partial }{\partial \xi}-
E^3\frac{\partial }{\partial z}\wedge\frac{\partial }{\partial \xi},
$$
$$
\bar{G}=\alpha_3\frac{\partial }{\partial x}\wedge\frac{\partial }{\partial y}-
\alpha_2\frac{\partial }{\partial x}\wedge\frac{\partial }{\partial z}+
\alpha_1\frac{\partial }{\partial y}\wedge\frac{\partial }{\partial z}+
B^1\frac{\partial }{\partial x}\wedge\frac{\partial }{\partial \xi}+
B^2\frac{\partial }{\partial y}\wedge\frac{\partial }{\partial \xi}+
B^3\frac{\partial }{\partial z}\wedge\frac{\partial }{\partial \xi}.
$$
The following relations hold:
$$
<F,\bar{F}>=<\beta,\mathbf{B}>-<\alpha,\mathbf{E}>, \ \ \
<G,\bar{G}>=<\alpha,\mathbf{E}>-<\beta,\mathbf{B}>,
$$
$$
<F,\bar{G}>=2<\alpha,\mathbf{B}>, \ \
<G,\bar{F}>=2<\beta,\mathbf{E}>,
$$
so, $<F,\bar{F}>=-<G,\bar{G}>$.

Now, since $\mathbf{d}\omega=0$, the zero value of the mentioned extension of
the Lie derivative (3) gives the equations
\begin{equation}
\mathcal{L}_{\bar{F}}\omega=\mathbf{d}i_{\bar{F}}\omega=-\mathbf{d}G=0,
\ \ \ \mathcal{L}_{\bar{G}}\omega=\mathbf{d}i_{\bar{G}}\omega=\mathbf{d}F=0 ,
\end{equation}
which reproduce Maxwell charge free equations (5) in a {\it metric-free way},
i.e., the two 2-forms $F,G$ are closed: $\mathbf{d}F=0, \ \mathbf{d}G=0$. In
other words, the flows of the bi-vectors $\bar{F}$ and $\bar{G}$ do NOT change
the volume 4-form $\omega$.

If $C^2_2$ is the contraction operator with
respect to the second member, then
the (1,1)-stress-energy-momentum tensor $\mathbb{T}_{(F,G)}$
admits the following coordinate-free and metric-free form
\begin{equation}
\mathbb{T}=-\frac12\left[C^2_2\,(F\otimes\bar{F})+C^2_2\,(G\otimes\bar{G})\right].
\end{equation}

Another look at $\mathbb{T}_{(F,G)}$ is the following.
Let $X$ be an arbitrary vector field on $\mathbb{R}^4$ and $\theta$ be an
arbitrary 1-form on $\mathbb{R}^4$. We can form now
$i_{X}F=X^\sigma
F_{\sigma\nu}dx^\nu$ and $i_{\theta}\bar{F}=
\theta_{\sigma}\bar{F}^{\sigma\nu}\frac{\partial }{\partial x^{\nu}}$.
The tensor $\mathbb{T}_{(F,G)}$ may now be defined in the following way:
$$
 \mathbb{T}(X,\theta)=
-\frac12\,tr\left[i_{X}F\otimes\,i_{\theta}\bar{F}+
i_{X}G\otimes\,i_{\theta}\bar{G}\right].
$$

In this way we get a relativistic {\it pre-metric} view on Maxwell
charge-free equations: no Minkowski pseudometric, no Hodge $"*"$.
(detailed comments see in [3]).

\section{Preliminary notes on non-linearization}
Nonlinearizations of Maxwell charge free electrodynamics the reader may find in
[4],...,[17] and [18],...,[21].

Passing to nonlinear field equations we begin with two appropriate examples
[21, p.370-371;22].

 \vskip 0.4cm
The examples consider the nonlinear equations defining the so called
{\it autoparallel time-like} and {\it null} vector fields with respect to a given
connection, restricting to the simple case of Levi-Civita connection
$\Gamma=(\Gamma_{\mu\nu}^\sigma)$
on Minkowski space-time $(\mathbb{R}^4,\eta)$.

Recalling the definition of autoparallel vector field: $\nabla_XX=0$, or
$i_{X}\nabla X=0$, it is interesting to note that this nonlinear system
of equations admits (3+1)-soliton-like, even {\it spatially finite}, solutions
on Minkowski space-time.

In fact, in canonical coordinates
$(x^1,x^2,x^3,x^4)=(x,y,z,\xi=ct)$ we have $\Gamma_{\mu\nu}^\sigma=0$, and let
 $u^\mu=(0,0,\pm \frac vc f,f)$,
be the components of the time-like vector field
$u, \ \eta(u,u)>0$, where $0<v=const<c$,
and $c$ is the velocity of light, so $\frac vc < 1$ and $\eta(u,u) =
\left(1-\frac{v^2}{c^2}\right)f^2>0$. Then every function $f$ of the kind $$
f(x,y,z,\xi)=f\left(x,y,\alpha.(z\mp\frac vc \xi)\right),\ \alpha=const,
\quad\text{for example}\quad \alpha=\frac{1}{\sqrt{1-\frac{v^2}{c^2}}}, $$
defines a solution.

If $\eta(u,u)=0$ then the equations
 are equivalent to $u^\mu(\mathbf{d}\tilde{\eta}(u))_{\mu\nu}=0$, where
$\mathbf{d}$ is the exterior derivative. In fact, since the connection used
is riemannian, we have $0=\nabla_\mu\frac12(u^\nu u_\nu)=u^\nu\nabla_\mu
u_\nu$, so the relation $u^\nu\nabla_\nu u_\mu -u^\nu\nabla_\mu u_\nu=0$
holds and is obviously equal to $u^\mu(\mathbf{d}u)_{\mu\nu}=0$. The
soliton-like solution is defined by $u^\mu=(0,0,\pm f,f)$ where the function $f$
is of the form
$$
f(x,y,z,\xi)=f(x,y,z\mp \xi).
$$
Clearly, for every autoparallel vector field $u$ (or one-form $u$) there
exists a canonical coordinate system on the Minkowski space-time, in
which $u$ takes such a simple form: $u^\mu=(0,0,\alpha f,f), \alpha=const$.
The dependence of $f$ on the three spatial coordinates $(x,y,z)$ is
{\it arbitrary} , so it is allowed to be chosen {\it soliton-like} and, even,
{\it spatially finite}.

So, although the trajectories of  these autoparallel vector fields are straight
lines and in this sense are naturally considered as appropriate description of
free point-like objects, the field nature of these vector fields deserves
corresponding attention and respect as a generator of (3+1)-spatially finite
solutions. Moreover, these equations suggest one possible way to find
appropriate nonlinearization in other more complex cases: having the
mathematical image $\Phi$ of the physical object of interest and its
change $\nabla\Phi$, compute the flow $\mathfrak{F}$ of $\Phi$ across
$\nabla\Phi$, and put this flow equal to zero. This means that the change
$\nabla\Phi$ is {\it admissible} for $\Phi$, i.e., the physical object
considered, appropriately changes under the action of external
fields of different nature in order to {\it survive}. Of course, if the object
considered consists of several interacting subsystems, then the external and
the mutual among the subsystems interaction has to be correspondingly identified
and formally represented in order to get the right entire and complete picture
of the object's appearance and behavior.

  In the frame of such approach we are going to nonlinearize Maxwell charge
free equations in a {\it pre-metric} way, trying to build spatially finite
image of a photon-like object, consisting of two interacting subsystems, and
the constituents of these two subsystems to be formally represented in
terms of already introduced four fields $(\mathbf{E},\mathbf{B};\alpha,\beta)$.

\vskip 0.3cm
The first step to nonlinearization considers the case, of course, of
{\it non-zero divergences} of $\mathbf{E}$ and $\mathbf{B}$, i.e.,
$\mathbf{d}i_{\mathbf{E}}\omega_3\neq 0, \ \
\mathbf{d}i_{\mathbf{B}}\omega_3\neq 0$, in general, and {\it available
local interaction} between the flows generated by the two vector fields
$(\mathbf{E},\mathbf{B})$, not ignoring, of course, the participation of their
1-form companions $(\alpha,\beta)$. Note that, in the linear case, the basic
constituents of the situation were these four objects, and no definite local
time-stable subsystems created by them were defined and used. Our view on
available intrinsic interaction, i.e., {\it local stress-energy-momentum
exchange}, inside an electromagnetic field object is based on the following
presumption: \vskip 0.3cm
{\bf Existence of recognizable time-stable subsystems of the object considered,
which subsystems are able to carry and exchange stress-energy-momentum, is
required}.
\vskip 0.3cm
Therefore, our first step should be to formally identify and further recognize
such subsystems. Our approach is based on the hypothesis that the subsystems
are just {\it two}. As {\it formal constituents} of these two
subsystems we choose the mathematical objects
$(\mathbf{E},\mathbf{B};\alpha,\beta)$, where, as we mentioned,
$(\mathbf{E},\mathbf{B})$ are allowed now to have NON-zero divergences, so, in
general,
$i_{\mathbf{E}}\mathbf{d}\alpha\neq 0$,
$i_{\mathbf{B}}\mathbf{d}\beta\neq 0$, $i_{\mathbf{E}}\mathbf{d}\beta\neq 0$ and
$i_{\mathbf{B}}\mathbf{d}\alpha\neq 0$.

\section{Pre-metric formulation in the non-linear case}
 \subsection{Static  case}
In this case in order to follow the above given view, we consider  defined on
$\mathbb{R}^3$ {\it vector-valued objects}, in terms of which the mathematical
images of {\it two subsystems} will be created. Since the subsystems are two,
our objects will take values in a 2-dimensional real vector space denoted
further by $V$, and equipped with a basis $(e_1,e_2)$.

The first subsystem $(\Omega,\bar{\Omega})$ consists of the following two
$V$-valued objects:
\begin{equation}
\Omega=\alpha\otimes e_1+\beta\otimes e_2, \ \ \
\bar{\Omega}=\mathbf{E}\otimes e_1+\mathbf{B}\otimes e_2.
\end{equation}
Recalling now that the two 1-forms $(\alpha,\beta)$ have their bi-vector
$\omega_3$-images $\bar{H}$ and $\bar{K}$ according to
$\alpha=i_{\bar{H}}\omega_3, \ \ \beta=i_{\bar{K}}\omega_3, \ \
$
we define the second subsystem $(\Sigma,\bar{\Sigma})$ as follows:
\begin{equation}
\Sigma=-i_{\mathbf{B}}\omega_3\otimes e_1+i_{\mathbf{E}}\omega_3\otimes e_2, \
\ \ \bar{\Sigma}=-\bar{K}\otimes e_1+\bar{H}\otimes e_2.
\end{equation}
Note that while $\Omega$ is $V$-valued 1-form, $\Sigma$ is $V$-valued 2-form.
Correspondingly, $\bar{\Omega}$ is $V$-valued vector field, and $\bar{\Sigma}$
is $V$-valued bi-vector field.

In this {\it static} situation, i.e., there are no running processes, any
admissible stress-energy exchange between the two subsystems should be realized
as {\it dynamical stress equilibrium}, so, it must have the
following special property: it must be {\it simultaneous} and {\it in equal
quantities}, otherwise the required static situation will be disturbed.
Formally this means that any local stress-energy {\it loss/gain} of
$(\Omega,\bar{\Omega})$ is simultaneously compensated by corresponding local
stress-energy {\it gain/loss} of $(\Sigma, \bar{\Sigma})$, and vice versa. We
could say that our two subsystems demonstrate {\it stable local stress
equilibrium}. These preliminary remarks suggest: such {\it
stress-equilibrium situations need NOT available nonzero interaction
stress-energy}, which corresponds to the Maxwell stress-energy tensor which
does not contain interaction stress-energy between the two constituents
$\mathbf{E}$ and $\mathbf{B}$: the full stress-energy is the sum of the
stress-energies carried by the two constituents:
$$
M^i_j(\mathbf{E},\mathbf{B})=M^i_j(\mathbf{E})+M^i_j(\mathbf{B})=
 \left(\mathbf{E}^i\mathbf{E}_j-
\frac12\mathbf{E}^2\delta^i_j\right)+
\left(\mathbf{B}^i\mathbf{B}_j-\frac12
\mathbf{B}^2\delta^i_j\right)
$$
$$
=\frac12\Big[\mathbf{E}^i\mathbf{E}_j+
\left(i_{\mathbf{E}}\omega_3\right)^{ik}\left(i_{\mathbf{E}}\omega_3\right)_{kj}
+\mathbf{B}^i\mathbf{B}_j+
\left(i_{\mathbf{B}}\omega_3\right)^{ik}\left(i_{\mathbf{B}}\omega_3\right)_{kj}\Big] .
$$

The corresponding equations, defining local stress equilibrium, must
represent the following picture: {\it each subsystem keeps locally its
stress-energy and the possible changes are mutually compensated}. Assuming that
these changes are {\it real}, their corresponding formal expressions {\it
should have tensor nature}, so, the most appropriate formal expressions seem to
be given by $\mathbf{d}\Omega$ and $\mathbf{d}\Sigma$. The corresponding flows
should also pay respect to the vector-valued nature of the formal images of the
object's subsystems. These specificities of the stress-energy exchange suggest:
the stress-energy balance between the two subsystems to make use of the
$"\vee"$-extension (2) of the interior product as follows:
 \begin{equation}
 i^{\vee}_{\bar{\Omega}}\mathbf{d}\Omega=
-i^{\vee}_{\bar{\Sigma}}\mathbf{d}\Sigma.
\end{equation}

Our view is that equation (11) adequately  corresponds to the fact that there
is NO interaction stress in $M^i_j(\mathbf{E},\mathbf{B})$: the whole stress
should be carried by $(\Omega,\bar{\Omega})$ and $(\Sigma,\bar{\Sigma})$, and
the hidden "dynamical" aspect of this equilibrium is adequately
represented by the $"\vee"$-extension of the interior product. Recalling that
$\mathbf{d}i_{X}\omega_3=(div\,X).\omega_3$ and the above notations:
$\alpha=i_{\bar{H}}\omega_3$ and $\beta=i_{\bar{K}}\omega_3$ we obtain

$$
\mathbf{d}\Omega=\mathbf{d}\alpha\otimes e_1+\mathbf{d}\beta\otimes e_2, \ \ \
\  \mathbf{d}\Sigma=-\mathbf{d}i_{\mathbf{B}}\omega_3\otimes
e_1+\mathbf{d}i_{\mathbf{E}}\omega_3\otimes e_2  ,
$$
$$
i^{\vee}_{\bar{\Omega}}\mathbf{d}\Omega=
i_{\mathbf{E}}\mathbf{d}\alpha\otimes e_1\vee e_1+
i_{\mathbf{B}}\mathbf{d}\beta\otimes e_2\vee e_2+
(i_{\mathbf{E}}\mathbf{d}\beta+
i_{\mathbf{B}}\mathbf{d}\alpha)\otimes e_1\vee e_2 ,
$$
$$
 i^{\vee}_{\bar{\Sigma}}\mathbf{d}\Sigma
=\beta.div(\mathbf{B})\otimes e_1\vee e_1+
\alpha.div(\mathbf{E})\otimes e_2\vee e_2-
[\beta.div{\mathbf{E}}+
\alpha.div{\mathbf{B}}]\otimes e_1\vee e_2 .
$$
 Hence, the balance relation (11) leads to
\begin{eqnarray*}
&&i_{\mathbf{E}}\mathbf{d}\alpha+div(\mathbf{B}).\beta=0 , \\
&&i_{\mathbf{B}}\mathbf{d}\beta+div(\mathbf{E}).\alpha=0, \\
&&i_{\mathbf{E}}\mathbf{d}\beta+i_{\mathbf{B}}\mathbf{d}\alpha-
div(\mathbf{E}).\beta-div(\mathbf{B}).\alpha=0 .
\end{eqnarray*}
The nonzero values of
$div\,\mathbf{E}$, $div\,\mathbf{B}$,
$i_{\mathbf{E}}\mathbf{d}\alpha$,
$i_{\mathbf{B}}\mathbf{d}\beta$, $i_{\mathbf{E}}\mathbf{d}\beta$ and
$i_{\mathbf{B}}\mathbf{d}\alpha$
provide possible interaction between the two subsystems
formally represent by $(\Omega,\bar{\Omega})$
and $(\Sigma,\bar{\Sigma})$, so, the two subsystems acquire status of
{\it interacting} subsystems of a larger {\it stress-balanced} system, i.e.,
being in the state of {\it dynamical equilibrium}.

 \vskip 0.3cm

\subsection{Time dependent case}

First we note that introducing {\it time} is considered here as a
{\it quantitative comparison of the courses of two physically independent
processes}, the one of which we call {\it referent}, e.g., the progress of
appropriate watch, then the other one attains significance of {\it
parametrized} process.

Hence, we have to specially note that the {\it time} parameter $t$ used in this
subsection we continue to consider as {\it external} to the spatial coordinates
$(x,y,z)$ parameter, and the corresponding referent process must NOT influence
the parametrized process. The main formal consequence of this consideration
is, as we mentioned earlier, that time-derivatives do {\it not} change the
tensor nature of the $t$-differentiated object.

Naturally, from physical viewpoint, any observed time change of the above
discussed stress balance in the {\it static} case should presume corresponding
{\it influence}, leading to its violation, and, of course, to violation of its
formal representation - relation (11). Physically it may be expected the
electromagnetic field object described to survive through some kind of {\it
time "pulsating"} at the space points, or through a {\it propagation as a
whole} in the 3-space, or, both. So: {\it the local static balance should be
replaced by an appropriate intrinsically compatible local dynamical and time
dependent balance}. Hence, in order to survive, our object must be able to
generate {\it appropriate} spatial changes inside any spatial area that
it occupies at any moment of its existence.

To this time-dependence of the behavior of our electromagnetic field object we
are going to give formal description by means of finding appropriate change of
the static balance equation (11).

Equation (11) formally postulates equivalence between two vector valued 1-forms,
so, any introduced influence object, representing how the  new time-dependent
balance would look like, is expected, formally, also to be vector valued 1-form,
containing appropriately first order $(\xi=ct)\,$-derivative(s) and valued in
the same vector space. This allows a natural return to the static balance
equation through setting this new object equal to zero.

Also, since the available spatial differential operators in (11) are just of
{\it first order}, it seems natural the corresponding formal influence object
to contain time derivatives of {\bf not higher} than first order. Clearly, in
view of the flow nature of the objects across their own spatially
differentiated objects in the static relation (11), the influence object is
expected to express formally also a flow, but a flow across {\it time
differentiated} object. Moreover, it should be expected also this time
dependence to generate direct mutual influence between the two now
time-dependent subsystems. Finally, since time derivation must not change the
tensor nature of the differentiated object, and since $\Omega$ is 1-form, then
the 2-form $\Sigma$ is the natural candidate to be $\xi$-differentiated, and
the $"\vee"$-flow of $\bar{\Omega}$ across the $\xi$-differentiated $\Sigma$
will give vector valued 1-form, which naturally
appears as appropriate formal measure of the local stress-energy time-exchange.
So, we may write
\begin{equation}
i^{\vee}_{\bar{\Omega}}\mathbf{d}\Omega+
i^{\vee}_{\bar{\Sigma}}\mathbf{d}\Sigma =
i^\vee_{\bar{\Omega}}\frac{\partial}{\partial \xi}\Sigma .
\end{equation}
This equation (12) gives the following three equations
$$
i_{\mathbf{E}}\mathbf{d}\alpha+(div\,\mathbf{B}).\beta=
-i_{\left(\frac{\partial \mathbf{B}}{\partial
\xi}\wedge\mathbf{E}\right)}\omega_3,
$$
$$ i_{\mathbf{B}}\mathbf{d}\beta+
(div\,\mathbf{E}).\alpha= i_{\left(\frac{\partial \mathbf{E}}{\partial
\xi}\wedge\mathbf{B}\right)}\omega_3
$$
$$
i_{\mathbf{E}}\mathbf{d}\beta+i_{\mathbf{B}}\mathbf{d}\alpha-
(div\,\mathbf{E}).\beta-(div\,\mathbf{B}).\alpha=
	i_{\left(\frac{\partial \mathbf{E}}{\partial
\xi}\wedge\mathbf{E}\right)}\omega_3 -
i_{\left(\frac{\partial \mathbf{B}}{\partial
\xi}\wedge\mathbf{B}\right)}\omega_3. $$ \vskip 0.3cm


\subsection{Space-time representation}
In the frame of the space-time view on physical processes the introduced
variable $\xi=ct$ is no more independent on the choice of physical frames with
respect to which we introduce spatial coordinates and write down time-dependent
formal relations. Now $\xi$ is considered as appropriate coordinate, it
generates local coordinate base vector $\frac{\partial}{\partial \xi}$ and
corresponding co-vector (or 1-form)
$d\xi, \langle d\xi,\frac{\partial}{\partial \xi}\rangle=1$.
So, the 3-volume $\omega_3=dx\wedge dy\wedge dz$
naturally extends to the 4-volume $\omega=dx\wedge dy\wedge dz\wedge d\xi$
on $\mathbb{R}^4$. Our purpose now is to find appropriate 4-dimensional balance
law, suggested by the previous balance laws formally given by equations
(11),(12).

 Recall our two basic objects: the vector valued differential 1-form
$\Omega=\alpha\otimes e_1+\beta\otimes e_2$ and the vector valued differential
2-form $\Sigma=-i_{\mathbf{B}}\omega_3\otimes e_1+
i_{\mathbf{E}}\omega_3\otimes e_2$, been defined entirely in
terms of objects previously introduced on $\mathbb{R}^3$. We want now
these objects to depend on $\xi$ as they depend on the spatial coordinates,
 so to be appropriately extended to objects on $\mathbb{R}^4$.

Now, the 4th dimension $\xi$ generates the coordinate 1-form $d\xi$, so,
the vector valued 1-form $\Omega$ turns to $d\xi$ for help to extend to a 2-form
on $\mathbb{R}^4$, which is done in the simplest way:
$\Omega\rightarrow\Omega\wedge d\xi$. We are in position now to consider the
difference $\Omega\wedge d\xi-\Sigma$.

$$
\Omega\wedge d\xi-\Sigma=(\alpha\wedge d\xi)\otimes e_1+
(\beta\wedge d\xi)\otimes e_2+i_{\mathbf{B}}\omega_3\otimes e_1-
i_{\mathbf{E}}\omega_3\otimes e_2
$$
$$
=(i_{\mathbf{B}}\omega_3+\alpha\wedge d\xi)\otimes e_1-
(i_{\mathbf{E}}\omega_3-\beta\wedge d\xi)\otimes e_2.
$$
In this way we get two differential 2-forms on $\mathbb{R}^4$ naturally
recognized by the basis vectors $(e_1,e_2)$ of the external vector space $V$:
 $$
F=i_{\mathbf{B}}\omega_3+\alpha\wedge d\xi \ \ \ \ \ \text{and} \ \ \ \
G=i_{\mathbf{E}}\omega_3-\beta\wedge d\xi .
 $$
These two 2-forms we consider further as vector components of {\it
one} $V$-valued 2-form $\mathbf{\Omega}$:
$$
 \mathbf{\Omega}=F\otimes e_1+G\otimes e_2.
$$

In order to define corresponding flow, as we did it in previous subsections,
we have to construct $\bar{\mathbf{\Omega}}$. The corresponding 2-vectors
$\bar{F}$ and $\bar{G}$ are easily introduced making use of the isomorphism
between 2-forms and 2-vectors defined by the volume 4-form
$\omega=dx\wedge dy\wedge dz \wedge d\xi$ according to
$$
G=-i_{\bar{F}}\omega, \ \ \ F=i_{\bar{G}}\omega: \ \ \rightarrow \ \ \
\bar{\mathbf{\Omega}}=\bar{F}\otimes e_1+\bar{G}\otimes e_2.
$$

 We turn now to the corresponding balance law, it reeds:
\begin{equation}
i^{\vee}_{\bar{\mathbf{\Omega}}}\mathbf{d}\mathbf{\Omega}=0,
\end{equation}
i.e., the $"\vee"$-flow of ${\bar{\mathbf{\Omega}}}$ across the change
$\mathbf{d}\mathbf{\Omega}$ of $\mathbf{\Omega}$ does NOT lead to losses. It has
to be noted,
 that this balance law is written down without making use of
(pseudo)metric, the volume form $\omega$ serves sufficiently well. We obtain:
$$
i^{\vee}_{\bar{\mathbf{\Omega}}}\mathbf{d}\mathbf{\Omega}=
i^{\vee}_{(\bar{F}\otimes
e_1+\bar{G}\otimes e_2)}(\mathbf{d}F\otimes e_1+ \mathbf{d}\mathbf{G}\otimes
e_2) $$ $$ =i_{\bar{F}}\mathbf{d}F\otimes e_1\vee e_1+
i_{\bar{G}}\mathbf{d}G\otimes e_2\vee e_2+
(i_{\bar{F}}\mathbf{d}G +i_{\bar{G}}\mathbf{d}F)\otimes e_1\vee e_2=0 .
$$

So, equation (13) gives the following three equations

\begin{equation}
 i_{\bar{F}}\mathbf{d}F=0, \ \ \
i_{\bar{G}}\mathbf{d}G=0, \ \ \
i_{\bar{F}}\mathbf{d}G+i_{\bar{G}}\mathbf{d}F=0.
\end{equation}
We have here maximum 12 equations for the 12 components of $(F,G)$.

We give now another form to equations (14) making use of the Poincare
isomorphisms $\mathfrak{D}$ and $\mathbb{D}$, recalling:
$\mathbb{D}_{n-p}\circ\mathfrak{D}^p=(-1)^{p(n-p)}id$ and
$\mathfrak{D}^{n-p}\circ\mathbb{D}_p=(-1)^{p(n-p)}id$,
also, the divergence operator
$\delta^p=(-1)^p\,\mathbb{D}_{n-p+1}\circ\mathbf{d}\circ \mathfrak{D}^p$.
Clearly, $\delta^p$ maps $p$-vector fields to $(p-1)$-vector fields. Since in
our case
$$
G=-i_{\bar{F}}\omega=-\mathfrak{D}\bar{F}, \ \ \
F=i_{\bar{G}}\omega=\mathfrak{D}\bar{G},
$$
where $\bar{F}$ and $\bar{G}$ are bi-vector fields, then $\delta\bar{F}$
and $\delta\bar{G}$ will be just vector fields. We obtain:
$$
\mathfrak{D}(\bar{F}\wedge\delta\bar{G})=i_{\delta\bar{G}}\mathfrak{D}(\bar{F})=
i_{\delta\bar{G}}(i_{\bar{F}}\omega)=i_{\delta\bar{G}}(-G)=
\mathfrak{D}(\delta\bar{G}\wedge\bar{F})=i_{\bar{F}}\mathfrak{D}\delta\bar{G}
$$
$$
=i_{\bar{F}}\mathfrak{D}^1(-1)^2\mathbb{D}_3\,\mathbf{d}\,\mathfrak{D}^2\bar{G}=
i_{\bar{F}}\,(-id)\mathbf{d}\,\mathfrak{D}^2\,\bar{G}=
-i_{\bar{F}}\mathbf{d}(i_{\bar{G}}\omega)=-i_{\bar{F}}\mathbf{d}F.
$$
So, $i_{\delta\bar{G}}G=i_{\bar{F}}\mathbf{d}F$. In the same way
following the same line of transformations, from
$\mathfrak{D}(\bar{F}\wedge\delta\bar{F})$,
$\mathfrak{D}(\bar{G}\wedge\delta\bar{F})$,
$\mathfrak{D}(\bar{G}\wedge\delta\bar{G})$
we obtain consecutively
$$
i_{\delta\bar{F}}F=i_{\bar{G}}\mathbf{d}G, \ \ \
i_{\delta\bar{F}}G=-i_{\bar{F}}\mathbf{d}G, \ \ \
i_{\delta\bar{G}}F=-i_{\bar{G}}\mathbf{d}F.
$$
 Hence, equations (14) are equivalent to
\begin{equation}
i_{\delta\bar{G}}G=0,\ \ \ i_{\delta\bar{F}}F=0,
\ \ \ i_{\delta\bar{F}}G+i_{\delta\bar{G}}F=0.
\end{equation}

Note that, from {\it algebraic} viewpoint, each of the two equations
$i_{\delta\bar{G}}G=0$, $i_{\delta\bar{F}}F=0$,
represents a homogeneous algebraic system of equations for the components of
$\delta\bar{G}$ and $\delta\bar{F}$, which components in the both cases are
just {\it four}. Since the {\it nonlinear} solutions would require
$\delta\bar{G}\neq 0$ and $\delta\bar{F}\neq 0$, a nonzero nonlinear solution
would be possible only if $det||(F_{\mu\nu})||=0$ and $det||(G_{\mu\nu})||=0$,
i.e., when $<\alpha,\mathbf{B}>= <\beta,\mathbf{E}>=0$, or, $F\wedge F=0,
G\wedge G=0$. In the Minkowski metric case this corresponds to
$F_{\mu\nu}(*F)^{\mu\nu}=0$, which means euclidean orthogonality of
$\mathbf{E}$ and $\mathbf{B}$ : $\mathbf{E}.\mathbf{B}=0$.

Summing up the above three equations (15) we obtain
$$
i_{(\delta\bar{F}+\delta\bar{G})}(F+G)=i_{\delta(\bar{F}+\bar{G})}(F+G)=0,
$$
and since in the nonlinear case $(\delta\bar{F}+\delta\bar{G})\neq 0$
in general, we must have $det||(F+G)_{\mu\nu}||=0$, i.e.,
$$
(F+G)\wedge(F+G)=F\wedge F+2F\wedge G+G\wedge G=2F\wedge G=
2(<\alpha,\mathbf{E}>-<\beta,\mathbf{B}>)=0.
$$
The two cases $\delta\bar{F}=\delta\bar{G}\neq 0$  and
$\delta\bar{F}=-\delta\bar{G}\neq 0$ also require
$<\alpha,\mathbf{E}>=<\beta,\mathbf{B}>$. In the Minkowski metric case
this is equivalent to $F_{\mu\nu}F^{\mu\nu}=0$, which in
the euclidean metric terms means $\mathbf{E}^2=\mathbf{B}^2$. Hence, in terms
of Minkowski spacetime we can say: all nonlinear solutions require {\it null}
nature of the bi-vector fields $\bar{F}$ and $\bar{G}$, as well as
{\it null} nature of the 2-forms $F=\mathfrak{D}\bar{G}$
and $G=-\mathfrak{D}\bar{F}$.

Finally, we may say that all {\it nonlinear} solutions of (14) require
$i^{\vee}_{\bar{\mathbf{\Omega}}}\mathbf{\Omega}=0$, which leads to \newline
$\mathcal{L}^{\vee}_{\bar{\mathbf{\Omega}}}\mathbf{\Omega}=0$.
 \vskip 0.3cm
The energy-momentum tensor is the same as in the linear relativistic case (8).
Here we approach the conservation aspects as follows. It should be clear that
the object considered {\it propagates}, so, every local conservation aspect
must be considered with respect to the vector field $X$, determining the
corresponding propagation. Therefore, the energy tensor (8) must be
appropriately projected along this vector field $X$, and the quantity under
consideration should look like $\mathbf{V}=X^\sigma
\mathbb{T}_{\sigma}^{\mu}\frac{\partial}{\partial x^{\mu}}$. Since this vector
field $\mathbf{V}$ must NOT lose, or gain, anything to/from the volume where it
propagates, the coordinate free adequate image to this property will require the
corresponding Lie derivative of the 4-volume form $\omega$ with respect
$\mathbf{V}$ to be zero: $L_{\mathbf{V}}\omega=0$. Now, since
$\mathbf{d}\omega=0$, for the Lie derivative we obtain
\begin{equation}
0=L_{\mathbf{V}}\omega=\mathbf{d}i_{\mathbf{V}}\omega+
i_{\mathbf{V}}\mathbf{d}\omega=\mathbf{d}i_{\mathbf{V}}\omega=
\mathbf{d}\mathfrak{D}(\mathbf{V}).
\end{equation}
The obtained 3-form $\mathfrak{D}(\mathbf{V})$ is closed, so the corresponding
to $X$ conserved quantity is explicitly represented, and restricting this
3-form to the 3-space, we can integrate over the occupied by the object 3-volume
in order to compute the integral value of this quantity, carried by the
electromagnetic object considered.
\vskip 0.3cm
It deserves now noting the following. In this coordinate free pre-metric
formulation we may forbid about how we came to the components of $F$ and $G$,
i.e., each of the new constituents $F$ and $G$ has six components and that's
all, getting back to $(\mathbf{E},\mathbf{B};\alpha,\beta)$ is not necessary.
\vskip 0.4cm
The equations obtained suggest some connection with the concepts of {\it
absolute} and {\it relative} integral invariants of a vector field $X$ on a
manifold $M$ introduced and used by E.Cartan [23]: these are differential forms
$\alpha\in \Lambda(M)$ satisfying respectively the relations $i(X)\alpha=0,
i(X)\mathbf{d}\alpha=0$, leading to $L_{X}\alpha=0$, and just
$i(X)\mathbf{d}\alpha=0$. Our relations may be considered as corresponding
extensions:

-{\it a vector field} $\rightarrow$ {\it vector valued multivector field}

-{\it a differential form} $\rightarrow$  {\it vector valued differential
form},

\noindent
and these extensions allow to make use of the mentioned in Sec.2 extension of
the Lie derivative of a differential form along {\it multivector} fields.

The new moment in our extension is that we consider vector valued multivectors
along which vector valued forms to be differentiated {\it with respect to some
bilinear map} $\varphi:V\times V\rightarrow W$, where $W$ is appropriately
determined vector space. In our case $\varphi=\vee$, which
corresponds to the {\it specific} kind of interaction between the two
subsystems: {\it absence of non-zero interaction stress-energy}.

\vskip 0.3cm In
general, we note that, the triple $(V,W;\varphi)$ determines possible
interactions among the subsystems of the field object considered. In our
considerations these subsystems are formally represented by the vector
components of the multi-vector $\mathbf{\bar\Omega}$ and the
vector components of the multi-differential form
$\mathbf{\Omega}$, i.e., $(\bar{F},F)$ and $(\bar{G},G)$.

\vskip 0.3cm

\section{Conclusion}
From general point of view,
getting knowledge for the internal compatibility and external stability of a
physical object is being done by measuring the corresponding to these physical
appearances appropriate physical quantities. Such physical quantities may vary
in admissible, or not admissible extent: in the first case we talk about
admissible changes, and in the second case we talk about changes leading to
destruction of the object. Formally, this is usually verified by calculating the
flow of the formal image of the (sub)system considered through its
appropriately modelled change, as it is seen, e.g., in (11),(12),(13), i.e., by
means of finding corresponding {\it differential self flows of the subsystems},
e.g.,  $i_{\bar{F}}\mathbf{d}F$,
and {\it differential mutual flows among the subsystems}, e.g.,
$i_{\bar{F}}\mathbf{d}G$. Since every
measuring process requires stress-energy-momentum transferring between the
object studied and the measuring system, the role of finding corresponding {\it
tensor} representatives of these change-objects and of the corresponding flows
is of serious importance. Therefore, having adequate stress-energy-momentum for
the considered case, the clearly individualized tensor members of its
divergence represent qualitatively and quantitatively important aspects of the
{\it intrinsic dynamical nature} of the object considered. This
view motivated the above given approach to find appropriate description of
electromagnetic field objects.

The existing knowledge about the structure and internal dynamics of free
electromagnetic field objects made us assume the notion for {\it two
partner-fields internal structure}, formally represented by
$(F,\bar{F};G,\bar{G})$ on $\mathbb{R}^4$.
 Each of these two partner-fields is built of two {\it
formal constituents}, and each partner-field is able
to carry local stress-energy-momentum, allowing internal local
"intercommunication" between its constituents during the local interaction with
its partner-field. The two subsystems carry equal local energy-momentum
densities, and realize local mutual energy exchange {\it without available
interaction energy}. Moreover, they strictly respect each other: the exchange
is {\it simultaneous} and in {\it equal quantities}, so, each of the two
partner-fields keeps its identity and recognizability. The corresponding
internal dynamical structure appropriately unifies translation and rotation
through unique space-time propagations as a whole with the fundamental
velocity. All linear solutions to (14) represent Maxwell charge-free solutions.
As it was shown [21], in the corresponding Minkowski space-time
consideration the new nonlinear solutions, i.e., those satisfying
$\mathbf{d}F\neq 0, \mathbf{d}*F\neq 0$, are {\it time-stable, they admit
FINITE SPATIAL SUPPORT, and demonstrate compatible translational-rotational
dynamical structure}.
\vskip 0.7cm
{\bf Appendix}
\vskip 0.3cm
A natural {\it formal} extension of the final formulation of our
 nonlinearization is the following.

 Consider an oriented $4n$-manifold $\mathcal{M}=(M^{4n},\omega)$. Denote
by $\Lambda^{2n}(\mathcal{M})$ the $2n$-differential forms  on $\mathcal{M}$,
and by $\mathfrak{X}^{2n}(\mathcal{M})$ the the antisymmetric
$2n$-vector fields on $\mathcal{M}$. Let now $(\bar{F}_1, \bar{F}_2, ...
,\bar{F}_n)$ be a set of lineary independent $2n$-vector fields
and $(\bar{G}_1,
\bar{G}_2, ... \bar{G}_{n})$ be another set of lineary independent
$2n$-vector fields, i.e., members of $\mathfrak{X}^{2n}(\mathcal{M})$.
Let $(i=1,2, ..., n)$ and $(j=n+1,n+2, ..., 2n)$. The Poincare
isomorphysms generate the corresponding $2n$-differential forms $(F^i, G^j)$  :
$$ F^i=i_{\bar{G}_{j}}\omega, \ \ \ G^j=-i_{\bar{F}_{i}}\omega, \ \ j=n+i. $$
Let $\mathbf{V}$ be a $2n$-dimensional real vector space with basis
$(e_1,e_2, ..., e_{2n})$, and
$\vee:\mathbf{V}\times\mathbf{V}\rightarrow (\mathbf{V}\vee\mathbf{V})$
denote the symmetrized tensor product.
We can now define the
$\mathbf{V}$-valued objects (summation along $i$ and $j$)
$$
\Omega=F^i\otimes e_i+G^j\otimes e_j, \ \ \
\bar{\Omega}=\bar{F}^i\otimes e_i+\bar{G}^j\otimes e_j ,
$$
and the $\mathbf{V}\vee\mathbf{V}$-valued {\it self-mutual} flow
$$
i^{\vee}_{\bar{\Omega}}\mathbf{d}\Omega =
i_{\bar{F}_{i}}\mathbf{d}F^i\otimes(e_i\vee e_i)+
i_{\bar{G}_{j}}\mathbf{d}G^j\otimes(e_j\vee e_j)+
(i_{\bar{F}_{i}}\mathbf{d}G^j+
i_{\bar{G}_{j}}\mathbf{d}F^i)\otimes(e_i\vee e_j).
 $$
So, if this {\it self-mutual} flow is zero :
$i^{\vee}_{\bar{\Omega}}\mathbf{d}\Omega = 0$, we obtain the equations
$$
i_{\bar{F}_{i}}\mathbf{d}F^i=0, \ \ i_{\bar{G}_{j}}\mathbf{d}G^j=0, \ \
i_{\bar{F}_{i}}\mathbf{d}G^j+i_{\bar{G}_{j}}\mathbf{d}F^i=0 ,
$$
meaning that every sub-object $(F^i,\bar{F}_{i})$ stays
recognizable during its interaction with all of $(G^j,\bar{G}_{j})$, and every
sub-object $(G^j,\bar{G}_{j})$ stays recognizable during its interaction with
all of $(F^i,\bar{F}_{i})$.

Finally we mention that if the sub-objects are {\it able to choose}
interaction partners inside this system, the corresponding partnership can be
specialized by replacing the symmetrized tensor product $"\vee"$ with another
{\it appropriate} bilinear map $\varphi: \mathbf{V}\times\mathbf{V}\rightarrow
\mathbf{W}$ to some vector space $\mathbf{W}$.

\newpage
\vskip 1.5cm
{\bf References}
\vskip 0.3cm

[1]. {\bf W.H. Greub},{\it Multilinear Algebra},
Springer Verlag, second edition, 1978, New York;
{\bf W.H. Greub, S.Halperin, R.Vanstone},{\it Connections, Curvature, and
Cohomology}, Vols. I-II, Academic Press, 1972-1973

[2]. {\bf W.M. Tulczyjew}, {\it The Graded Lie Algebra of Multivector Fields
and the Generalized Lie Derivative of Forms}, Bull. Acad. Pol. Sci. SMAP 22
(1974) 937-942; {\it The Poisson Bracket for Poisson Forms in Multisymplectic
Field Theory}, arXiv: math-ph/0202043v1

[3]. {\bf D.A.Delphenich}, http://www.neo-classical-physics.info/electromagnetism-and-
wave-theory.html, "{\it Pre-metric Electromagnetism}", parts I, II, e-book.

[4]. {\bf M. Born, L.Infeld}, {\it Proc.Roy.Soc.}, {\bf A 144}, 425 (1934)

[5]. {\bf W. Heisenberg, H. Euler}, {\it Zeit.Phys.}, {\bf 98}, 714 (1936)

[6]. {\bf M. Born}, {\it Ann. Inst. Henri Poincare}, {\bf 7}, 155-265 (1937).

[7]. {\bf J. Schwinger}, {\it Phys.Rev}. ,{\bf 82}, 664 (1951).

[8]. {\bf H. Schiff}, {\it Proc.Roy.Soc.} {\bf A 269}, 277 (1962).

[9]. {\bf J. Plebanski}, {\it Lectures on Nonlinear Electrodynamics}, NORDITA,
Copenhagen, 1970.

[10]. {\bf G. Boillat}, {\it Nonlinear Electrodynamics: Lagrangians and
Equations of Motion}, \newline J.Math.Phys. {\bf 11}, 941 (1970).

[11]. {\bf B. Lehnert, S. Roy}, {\it Extended Electromagnetic Theory}, World
Scientific, 1998.

[12]. {\bf D.A. Delphenich}, {\it Nonlinear Electrodynamics and QED},
arXiv:hep-th/0309108, (good review article).

[13]. {\bf B. Lehnert}, {\it A Revised Electromagnetic Theory with Fundamental
Applications}, Swedish Physic Arhive, 2008.

[14]. {\bf D. Funaro}, {\it Electromagnetsm and the Structure of Matter},
Worldscientific, 2008; also: {\it From photons to atoms}, arXiv: gen-ph/1206.3110
(2012).

[15]. {\bf E. Schrodinger}, {\it Contribution to Born's new theory of
electromagnetic feld}, Proc. Roy. Soc. Lond. {\bf A 150}, 465 (1935).

[16]. {\bf G. Gibbons, D. Rasheed}, {\it Electric-magnetic duality rotations in
non-linear electrodynamics}, Nucl. Phys. {\bf B 454} 185 (1995) hep-th/9506035.

[17] {\bf R. Kerner, A.L. Barbosa, D.V. Gal'tsov}, {\it Topics in Born-Infeld
Electrodynamics}, arXiv: hep-th/0108026 v2

[18]. {\bf S.G.Donev}, {\it A particular nonlinear generalization of Maxwell
equations admitting spatially localized wave solutions},
Compt.Rend.Bulg.Acad.Sci., vol.34, No.4 (1986).

[19]. {\bf S. Donev, M. Tashkova}, {\it Energy-momentum directed
nonlinearization of Maxwell's pure field equations}, Proc.R.Soc.Lond. A ,
1993, {\bf 443}, 301-312.

[20]. {\bf S. Donev, M. Tashkova},
{\it Energy-Momentum Directed Nonlinearization of Maxwell's  Equations in the
Case of a Continuous Medium} /Donev, S., Tashkova, M./,   Proc.R.Soc. Lond.A
{\bf 450}, 281 (1995)

[21]. {\bf S. Donev, M. Tashkova}, {\it Geometric View on Photon-like Objects},
 LAMBERT Academic Publishing, 2014 (also: arXiv,math-ph, 1210.8323v2)

[22]. {\bf S. Donev}, {\it Geodesic  Vector Fields on  Minkowski Space-Time
and (3+1)-Solitary Waves}, Commun.JINR - Dubna, E2-88-107.

[23]. {\bf E. Cartan}, {\it Lecons sur les invariants integraux}.
Cours professe a la Faculte des sciences de Paris, 1920-1921.

 \end{document}